\documentclass[twocolumn]{aastex62}
\usepackage{natbib}
\newcommand{\citeg}[1]{\citep[e.g.,][]{#1}}
\newcommand{\citef}[2]{\citep[#2;][]{#1}}
\newcommand{\citewt}[2]{\citep[#2][]{#1}}

\usepackage{multirow}
\usepackage{comment}
\usepackage{graphicx}
\usepackage{amsmath}
\usepackage[flushleft]{threeparttable}
\usepackage{dcolumn}

\def\JB{{\rm Jy~beam^{-1}}}
\def\mJB{{\rm mJy~beam^{-1}}}

\def\kms{{\rm km~s^{-1}}}

\def\deg{$^{\circ}\ $}

\graphicspath{{./}{figures/}}

\accepted{Nov 9, 2020}

\submitjournal{ApJ}

\shorttitle{Multi-epoch observations of the L1448C(N) protostellar jet}
\shortauthors{Yoshida et al.}

\begin{document}

\title{Multi-epoch SMA observations of the L1448C(N) protostellar SiO jet}

\correspondingauthor{Tomohiro Yoshida}
\email{yoshida.tomohiro.45z@st.kyoto-u.ac.jp}

\author[0000-0001-8002-8473]{Tomohiro Yoshida}
\affil{Department of Astronomy, Kyoto University, Kitashirakawa-Oiwake-cho, Sakyo-ku, Kyoto, 606-8502, Japan}

\author[0000-0000-0000-0000]{Tien-Hao Hsieh}
\affil{Academia Sinica Institute of Astronomy and Astrophysics, 11F of ASMA Building, No.1, Sec. 4, Roosevelt Rd, Taipei 10617, Taiwan}

\author[0000-0001-9304-7884]{Naomi Hirano}
\affil{Academia Sinica Institute of Astronomy and Astrophysics, 11F of ASMA Building, No.1, Sec. 4, Roosevelt Rd, Taipei 10617, Taiwan}

\author[0000-0002-8238-7709]{Yusuke Aso}
\affil{Korea Astronomy and Space Science Institute (KASI), 776 Daedeokdae-ro, Yuseong-gu, Daejeon 34055, Republic of Korea}



\begin{abstract}
L1448C(N) is a young protostar in Perseus, driving an outflow and an extremely high-velocity (EHV) molecular jet.
We present multi-epoch observations of SiO $J = 8-7$, CO $J = 3-2$ lines, and 345 GHz dust continuum toward L1448C(N) in 2006, 2010, and 2017 with the Submillimeter Array. 
The knots traced by the SiO line show the averaged proper motion is $\sim0\farcs06~{\rm yr^{-1}}$ and $\sim0\farcs04~{\rm yr^{-1}}$ for the blue- and red-shifted jet, respectively.
The corresponding transverse velocities are $\sim78~\kms$ (blueshifted) and $\sim52~\kms$ (redshifted).
Together with the radial velocity, we found the inclination angle of the jets from the plane of the sky to be $\sim34$\deg for the blueshifted jet and $\sim46$\deg for the redfshifted jet.
Given the new inclination angles, the mass-loss rate and mechanical power were refined to be $\sim1.8\times 10^{-6}~M_\odot$ and $\sim1.3~L_\odot$, respectively.
In the epoch of 2017, a new knot is detected at the base of the redshifted jet.
We found that the mass-loss rate of the new knot is three times higher than the averaged mass-loss rate of the redshifted jet.
Besides, continuum flux has enhanced by $\sim37\%$ between 2010 and 2017.
These imply that the variation of the mass-accretion rate by a factor of $\sim3$ has occurred in a short timescale of $\sim10-20$ yr.
In addition, a knot in the downstream of the redshifted jet is found to be dimming over the three epochs.
\end{abstract}


\section{Introduction} \label{sec:intro}
Studying protostellar jets/outflows helps us to understand the star formation process.
Some protostars at an early evolutionary stage have extremely high-velocity (EHV) jets \citeg{bac96}, which may play an important role for removing angular momentum and thus allowing circumstellar material to accrete onto the central stars.
In addition, molecular jets are believed to be launched from the innermost region of an accretion disk. Although the launching mechanism is still unclear, disk accretion should provide the material to be ejected in the form of molecular jets. Thus, studies of the jet properties provide us a chance to explore the collapse and accretion processes in star formation \citewt{lee20}{see the review by}.
Interestingly, most molecular jets have knotty structure \citeg{lee20}. The chain of knots not only indicates the periodic ejection of the jet \citeg{vor18} but can also be a useful tracer of the jet motion on the plane of the sky \citeg{gir01,jha15}.

L1448C (or L1448-mm) in the Perseus molecular complex \citef{ort18}{$d=293$ pc} is a good example of an outflow-driving source with EHV jets. The EHV components have been observed in several SiO transitions \citeg{bac91,nis07}.
Such SiO detections suggest molecular formation via gas-phase reactions by a high mass-loss rate ($\sim10^{-6}~M_\odot~{\rm yr^{-1}}$) jet launched from the inside of the dust sublimation radius of the protostar \citeg{gla91,tab17,lee20}.

A high angular resolution ($\sim$1$\arcsec$) study in L1448C with the Submillimeter Array \citef{ho04}{SMA} was performed by \citet{hir10}.
L1448C consists of two sources, L1448C(N) and L1448C(S), for which L1448C(N) is found to power the outflows with EHV jets.
The two-sided jet consists of a chain of knots that are deflected at $\sim10\arcsec$ from the central source.
The authors attributed this deflection to the precession or wobbling of the disk in an unresolved binary system.
Furthermore, the direction of the disk-jet system shows the oscillation could also occur in a single star system if the rotation axis is inclined to the global magnetic field \citep{mac20}.
The physical parameters of the outflow were derived including outflow mass, momentum, kinetic energy, mean velocity, dynamical timescale, mass-loss rate, momentum supply rate, and mechanical power.
As a result, \citet{hir10} suggested that the mass and the age of the central star are $0.03-0.09~M_\odot$ and $(4-12)\times10^{3}$ yr, respectively, implying that the central star is in an extremely early stage of evolution.

However, some of the physical parameters in \citet{hir10} are inclination dependent. To derive the inclination angle, \citet{gir01} observed the proper motion of the SiO clumps together with the radial velocity by comparing the observations with the Plateau de Bure Interferometer (PdBI) in 1990 and the BIMA millimeter array in $1998-1999$.
They suggested that the inclination angle of the jets with respect to the plane of the sky is $\sim$21\deg. However, their analysis was done using observations with a low angular resolution of $\sim$3$\arcsec$ which is much larger than the sizes of the jet knots ($\sim$1$\arcsec$) resolved in the SMA map.
Besides, the emission from the jet knots is extended in the lower excitation line (SiO $J=2-1$), carrying a large positional uncertainty of the jet knots.
Furthermore, the observations for the two epochs can introduce an uncertainty to the inclination angle given the different resolutions and sensitivities.

In this study, we present 3 epoch SiO $J=8-7$ and CO $J=3-2$ observations of L1448C(N) with the SMA from 2006 to 2017. The angular resolution of $\sim$1$\arcsec$ better identifies the proper motions of the individual knots. In addition, the higher excitation state of SiO $J=8-7$ can trace the shock more precisely.
In section \ref{sec:obs}, we describe the observational information. In section \ref{sec:res}, we report the results of our analysis; positional shift of the central source, proper motions and transverse velocities of the jet knots, inclination angles of the jets, physical parameters of the jets and outflows, appearance of a new knot, dimming of the downstream knots, and variability of the central source. We discuss the results in section \ref{sec:dis}, and we conclude in section \ref{sec:con}.

\section{Observations} \label{sec:obs}

The observations of 3 epochs were carried out with the SMA in 2006, 2010, and 2017. The observation dates, the array configurations, the synthetic beam sizes, and the rms noise levels at $1.0~\kms$ width are shown in Table \ref{tab:obs}. All observations contain the SiO $J=8-7$ line (347 GHz) and CO $J=3-2$ line (346 GHz). 
The SiO and CO lines were observed in the lower sideband in 2006 and 2017, but in the upper sideband in 2010.

The primary beam of the SMA has a size of $\sim35\arcsec$ at 345 GHz.
In 2006, two pointings separated by $17\arcsec$ were observed, while in 2010 and 2017, only single pointing centered at the position of L1448C(N) ($\alpha$(J2000) = 3$^{\rm h}$25$^{\rm m}$ 38$^{\rm s}$.87, $\delta$(J2000) = 30$^{\rm d}$44$^{\rm m}$05$^{\rm s}$.35) was observed. 

The data in 2006 have been reported in \citet{hir10}, and here we discuss the observations in 2010 and 2017.
The data in 2010 consist of extended and very extended array configurations (Table \ref{tab:obs}).
The data in 2010 were obtained using the ASIC correlator, which divided each sideband of 4 GHz bandwidth into 48 ``chunks" of 104 MHz width.
We used the configuration that gave 256 channels per chunk for the SiO line and 128 channels per chunk for the CO line.
The corresponding velocity resolution were 0.35 km s$^{-1}$ for the SiO line and 0.7 km s$^{-1}$ for the CO line.
The data were calibrated with the MIR package \citep{qi05}.
The nearby quasar 3C84 was used for amplitude and phase calibrations.
The flux calibrators were Neptune in the very extended configuration, and Callisto and Ganymede in the extended configuration.
The bandpass was calibrated by observing 3C454.3.
The calibrated visibility data were Fourier tansformed and CLEANed using the MIRIAD package \citep{sau95}.
The image cubes of SiO and CO were made with a velocity interval of 1 km s$^{-1}$.
The synthesized beam size of the SiO map was 0\farcs58$\times$0\farcs41 with a position angle of 82$^{\circ}$ and that of the CO map was 0\farcs58$\times$0\farcs41 with a position angle of 84$^{\circ}$ with a robust weighting of 0.5.

Observations of the 3rd epoch in 2017 were made in the extended configuration using the SWARM correlator that covers 8 GHz bandwidth with 4 ``chunks" of 2 GHz width and two orthogonally polarized receivers, Rx345 and Rx400, simultaneously.
Two receivers were tuned to the same frequency setting, which covered the frequency ranges of 340.3--348.3 GHz and 356.3--364.3 GHz in the lower sideband and upper sideband, respectively.
The spectral resolution is 140 kHz across the entire band.
The velocity resolution corresponds to 0.12 km s$^{-1}$ at 347 GHz.
The data of each receiver were calibrated independently using the MIR package \citep{qi05}.
The quasars 3C84 and 3C273, and Uranus were used as the gain calibrator, the bandpass calibrator and the flux calibrator, respectively. Then, the calibrated data were Fourier transformed and CLEANed using the Common Astronomy Software Application \citef{mcm07}{CASA} version 5.4 for the SiO line, the CO line, and the 345 GHz continuum emission.
In order to improve the sensitivity, the data of two receivers were combined when making the images.

The primary beam correction was applied to the images of the 2nd and 3rd epochs.
In order to fairly compare these images between the 3 epochs, the SiO and CO images of the 2nd and 3rd epoch observations with higher resolutions were re-gridded to the same cell size and convolved to the same beam as those of the 2006 observations with the lowest resolution.
The continuum images are made using the data with the {\it uv} distance of $>$ 70 $k{\lambda}$ for all the three epochs.
The continuum maps in 2006 and 2010 were imaged by combining the lower and upper sideband. On the other hand, in 2017, only the lower sideband was used because of wider bandwidth for the new receiver (8 GHz).
Then, the images were re-grided and convolved to the same cell size and beam size as those of the 2017 observations.
The resulting beam size of the continuum image was 0\farcs79$\times$0\farcs63 with a position angle of $-85^{\circ}$.
After the convolution, the rms noise level of the 345 GHz continuum emission was 6.0 $\mJB$.

\begin{center}
\begin{deluxetable*}{lcccc}

\tablecolumns{5}
\tablewidth{0pc}
\tablecaption{Properties of each observation}

\tablehead{
\colhead{Date} & \colhead{Array configuration} & \colhead{Beam size} & \colhead{Weighting} & \colhead{Noise level (K)\tablenotemark{a}}}
\startdata
\hline
12/5/2006 & Compact & \multirow{2}{*}{$0\farcs96 \times 0\farcs84$} &  \multirow{2}{*}{uniform} & \multirow{2}{*}{1.9} \\
12/25/2006 & Extended &  \\ \hline
9/17/2010 & Extended & \multirow{2}{*}{$0\farcs58 \times 0\farcs41$} & \multirow{2}{*}{robust=0.5} & \multirow{2}{*}{1.5}\\
7/19/2010 & Very Extended & \\ \hline
11/10/2017 & Extended & $0\farcs83 \times 0\farcs69$ & robust=0.5 & 2.3 \\
\hline
\enddata
\tablenotetext{a}{The rms noise levels of the SiO ($J = 8-7$) maps at the channel width of $1~\kms$ measured after being concolved to the same beam.}
\end{deluxetable*}
\label{tab:obs}
\end{center}

\section{Results} \label{sec:res}

\subsection{Shift of the continuum peak} \label{sec:propcont}

\begin{figure*}
\epsscale{1.0}
\plotone{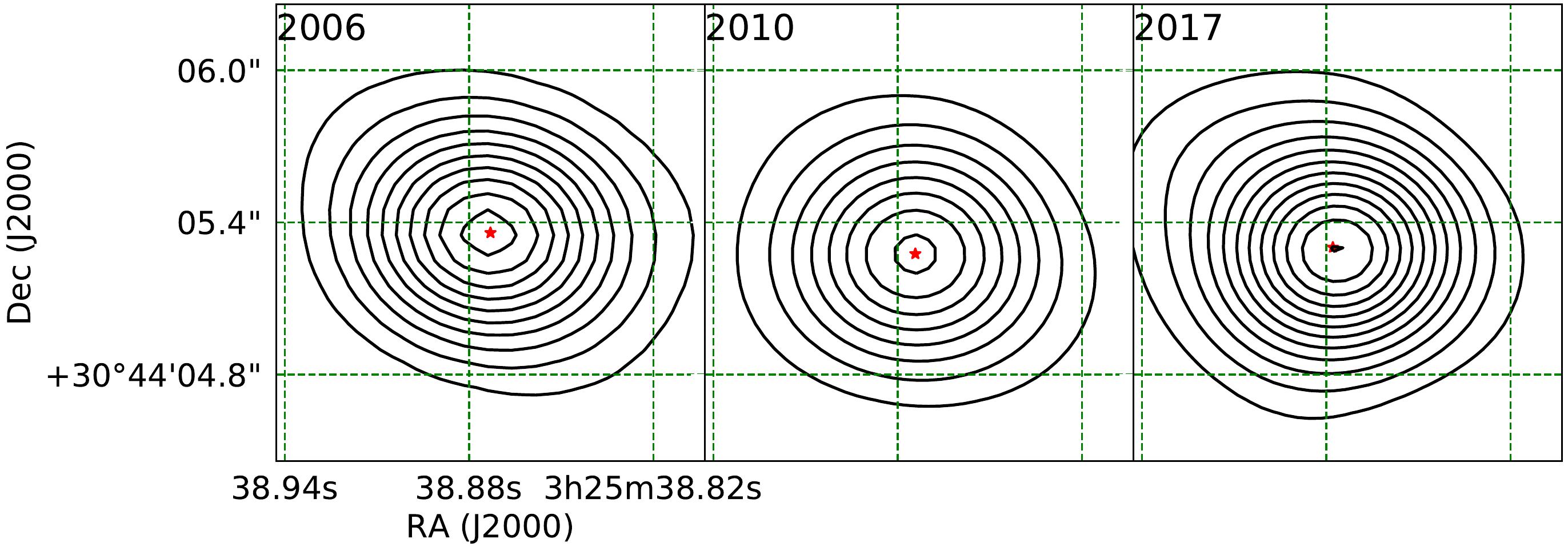}
\caption{Continuum maps made using the visibility data with the {\it uv} distance $> 70~{\rm k\lambda}$ in three epochs. The first contour is 5$\sigma$ and the contour interval is 5$\sigma$ where 1$\sigma =$ 6.0 $\mJB$. The red stars indicate the central positions obtained by fitting a Gaussian. The green grids are the same in all panels. \label{fig:propcont}}
\end{figure*}

In order to study the SiO proper motion, we firstly checked the continuum source positions in the three epochs. Figure \ref{fig:propcont} shows the continuum images and their central positions in the 3 epoch observations. 
The continuum centers are obtained by fitting a 2D-Gaussian in the image domain.

\begin{table}
\begin{center}

\caption{Central positions of the continuum emission}
\label{tab:cent}
\begin{tabular}{lcc}
\hline \hline
Epoch & \multicolumn{2}{c}{Central position (J2000)} \\
\hline
2006 & $3^{\rm h}25^{\rm m}38.873\pm0.00051$ & $30^{\rm d}44^{\rm m}05.36\pm0.0042$ \\
2010 & $3^{\rm h}25^{\rm m}38.874\pm0.00038$ & $30^{\rm d}44^{\rm m}05.28\pm0.0033$ \\
2017 & $3^{\rm h}25^{\rm m}38.878\pm0.00072$ & $30^{\rm d}44^{\rm m}05.30\pm0.0065$ \\
\hline
\end{tabular}
\end{center}
\end{table}

The central positions and their uncertainties from the fitting are listed in Table \ref{tab:cent}. The amounts of shift of the central positions are much smaller than the beam size. The uncertainties of the fitting were up to $\sim$11 mas in the direction of R.A., and $\sim$6.5 mas in the direction of Dec. Thus, the central positions of the source were measured well, and are used as the reference position for the jet knots (section \ref{sec:propsio}).

The amount of motion was estimated by averaging the shifts between two adjacent epochs.
The proper motion of the central source was estimated to be $\sim0\farcs012~{\rm yr^{-1}}$ in the direction from northwest to southeast, with a position angle of $\sim120^\circ$.

\subsection{Proper Motions of the SiO knots}\label{sec:propsio}

\begin{figure}
\epsscale{1.1}
\plotone{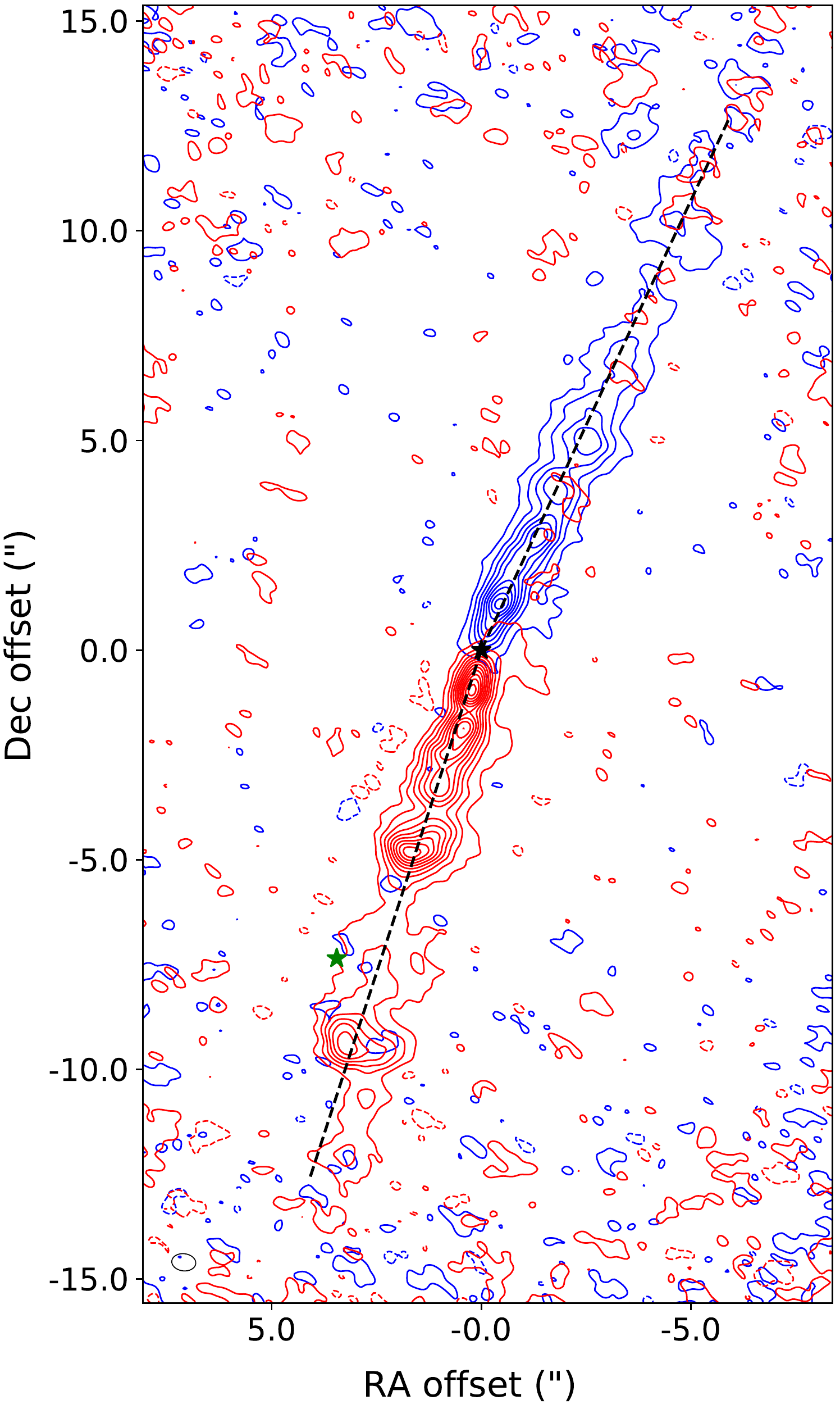}
\caption{High-velocity SiO emission observed in 2010 with the original angular resolution. The velocity ranges are $\Delta V = \pm41-70~\kms$ with respect to the systemic velocity ($v_{\rm LSR} \sim 5~\kms$). The contours start at 3$\sigma$ and increase in 5$\sigma$ intervals with 1$\sigma=0.23~{\rm Jy~beam^{-1}}$. The black broken lines indicate the PV cuts with position angle of $\sim335^\circ$ for the blueshifted part and $\sim162^\circ$ for the redshifted part. The black and green stars indicate the positions of the protostars, L1448C(N) and L1448C(S), respectively.\label{fig:mom0}}
\end{figure}

\begin{figure}
\epsscale{1.1}
\plotone{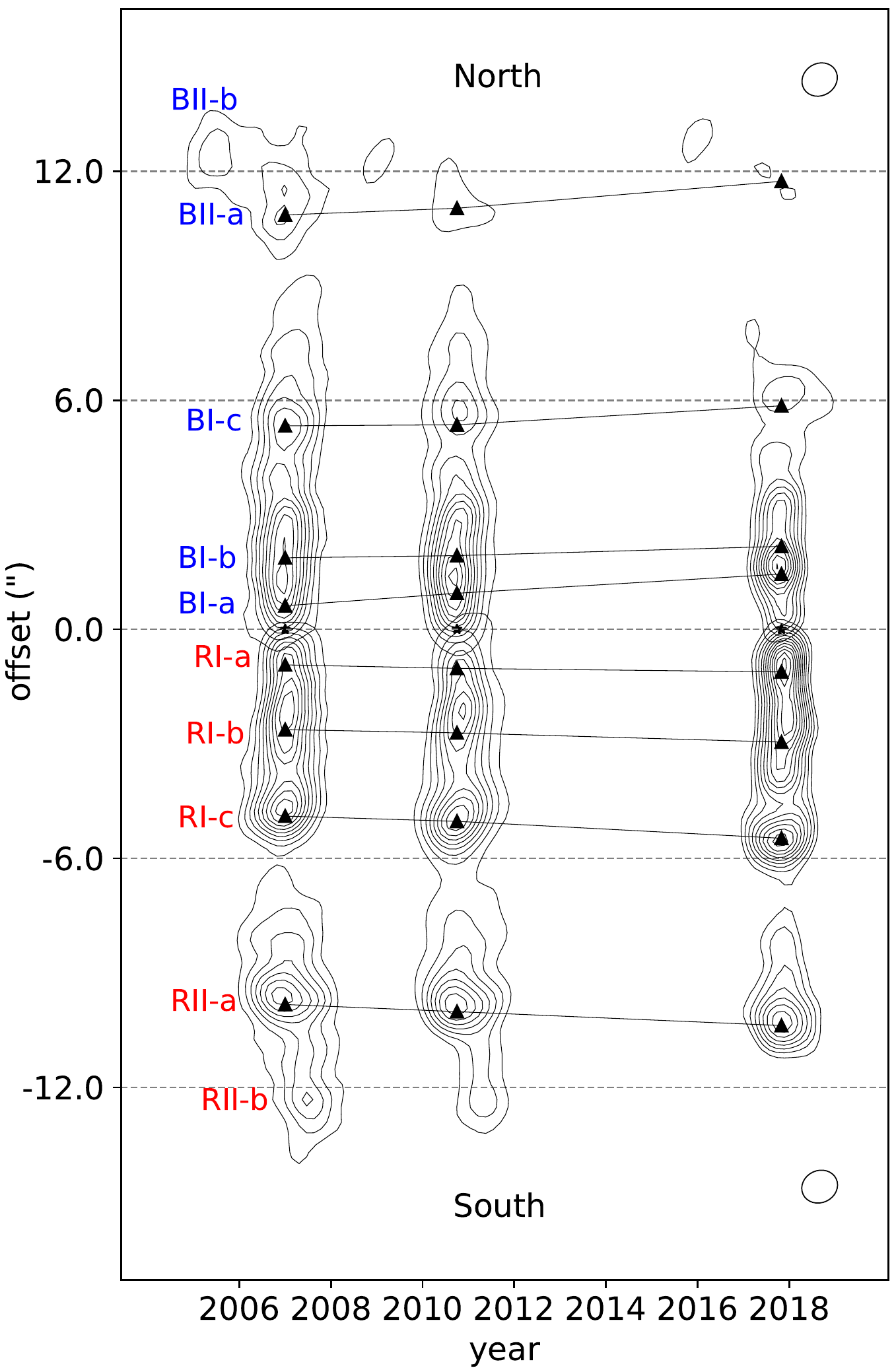}
\caption{Comparison of the SiO moment 0 maps ($\Delta V = \pm41-70~\kms$) in 3 epochs. The images of the blue- and redshifted jets were rotated counterclockwise by 25$^{\circ}$ and 18$^{\circ}$, respectively, so that the jet axes align with the vertical line.
The contours start at 3$\sigma$ and increase in 3$\sigma$ intervals with 1$\sigma=0.73~{\rm Jy~beam^{-1}}$.
The ellipses on the right top and bottom corner are the synthesized beams for the blue- and red-shifted side, respectively. The triangles are the peak positions of each knot in the PV diagrams (Figure \ref{fig:pv}). BI-a/RI-a and BI-b/RI-b are not separated in the moment 0 map, while they are different knots in Figure \ref{fig:pv}. \label{fig:comp}}
\end{figure}

High-velocity SiO emission observed in 2010 is shown in Figure \ref{fig:mom0}.
Note that Figure \ref{fig:mom0} shows the original resolution image before the convolution and re-griding.
The SiO knots are found in the EHV jets ($>40~\kms$) from L1448C(N).
The position angle of the EHV jet is ${\sim}{-}$25$^{\circ}$ in the blueshifted component, while it is ${\sim}{-}$18$^{\circ}$ in the redshifted component.
With the high angular resolution, the SiO jet is resolved in its transverse direction especially in the redshifted part.
The width of the jet increases gradually toward the downstream with an opening angle of $\sim$10$^{\circ}$. 
The intrinsic width of the jet (deconvolved by the beam) is $\sim0\farcs28$ (85 au) at a projected distance of 300 au from the driving source, and becomes $\sim1\farcs3$ (380 au) at a projected distance of 3000 au.

Figure \ref{fig:comp} compares the EHV jets at the three epochs.
The images are rotated by 25$^\circ$ and 18$^\circ$ for the blue- and red-shifted components, respectively so that the jets are aligned along the vertical axis.
The triangles in Figure \ref{fig:comp} indicate the knot positions that were obtained in the position-velocity diagrams (see the next paragraph). 
The outermost knots, BII-b and RII-b, were not included in this analysis because those knots are misaligned with the jet axis.

\begin{figure*}
\epsscale{0.9}
\plotone{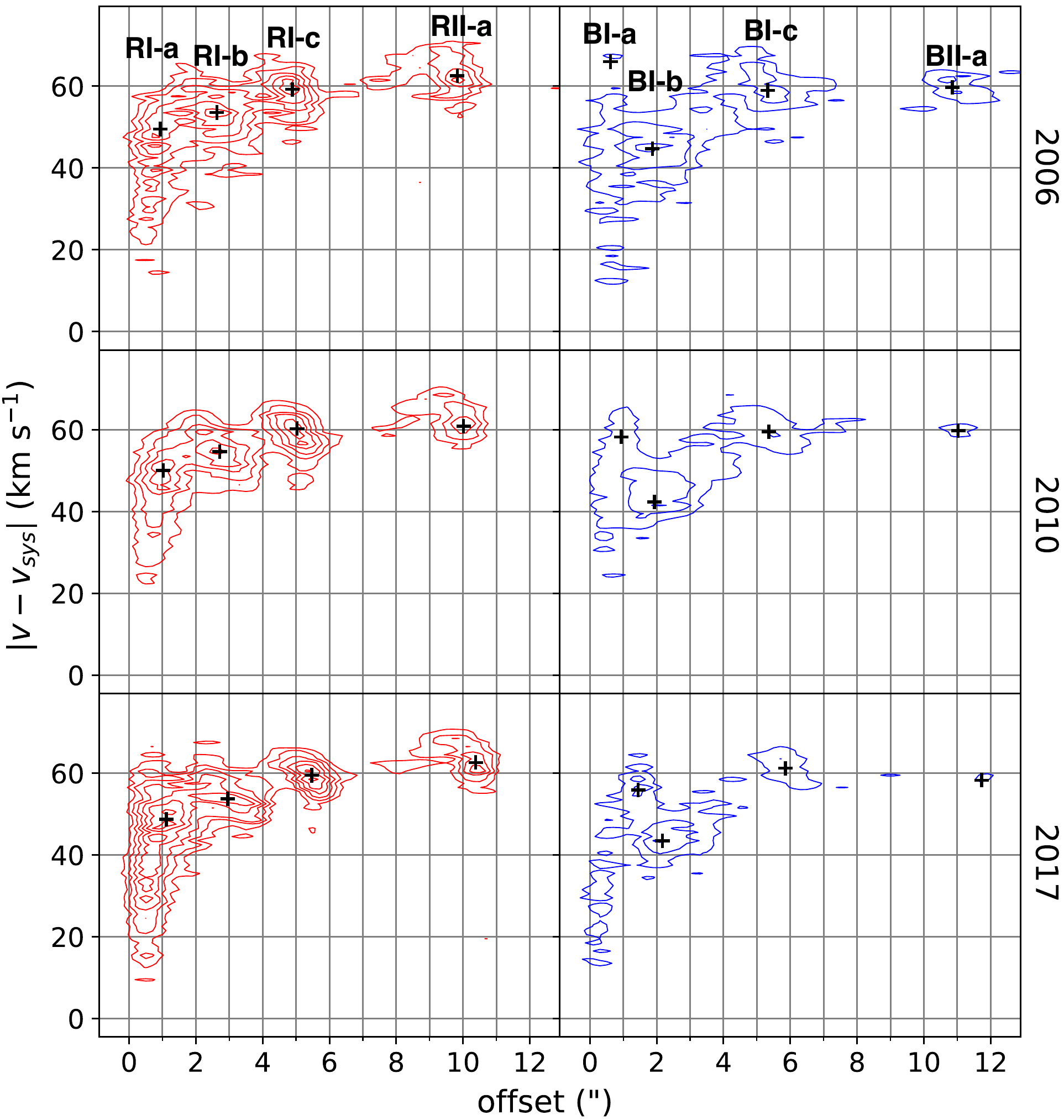}
\caption{PV diagrams of the SiO jets in the 3 epochs. The contours start at 2$\sigma$ and increase in 3$\sigma$ intervals where $1\sigma=0.13~\JB$. The positions of each knot are marked with black crosses. The systemic velocity of this source, $v_{\rm sys}$, is $\sim5~\kms$. \label{fig:pv}}
\end{figure*}

To obtain the positions and radial velocities of each knot, position velocity (PV) diagrams were used.
Figure \ref{fig:pv} shows the PV diagrams of these 3 epochs along the SiO jet axis.
The PV cuts are shown with the black broken lines in Figure \ref{fig:mom0}. The origins of the jets in each epoch were adopted to be the central position of the continuum emission described in section \ref{sec:propcont}.
In the high velocity part of this diagram (faster than $\pm40~{\rm km~s}^{-1}$), several peaks were found.
We identified these knots in the PV diagram by 2D-hyperboloid fitting.
Table \ref{tab:knots} shows the central position and the radial velocity of each knot.

\begin{figure}
\epsscale{1.2}
\plotone{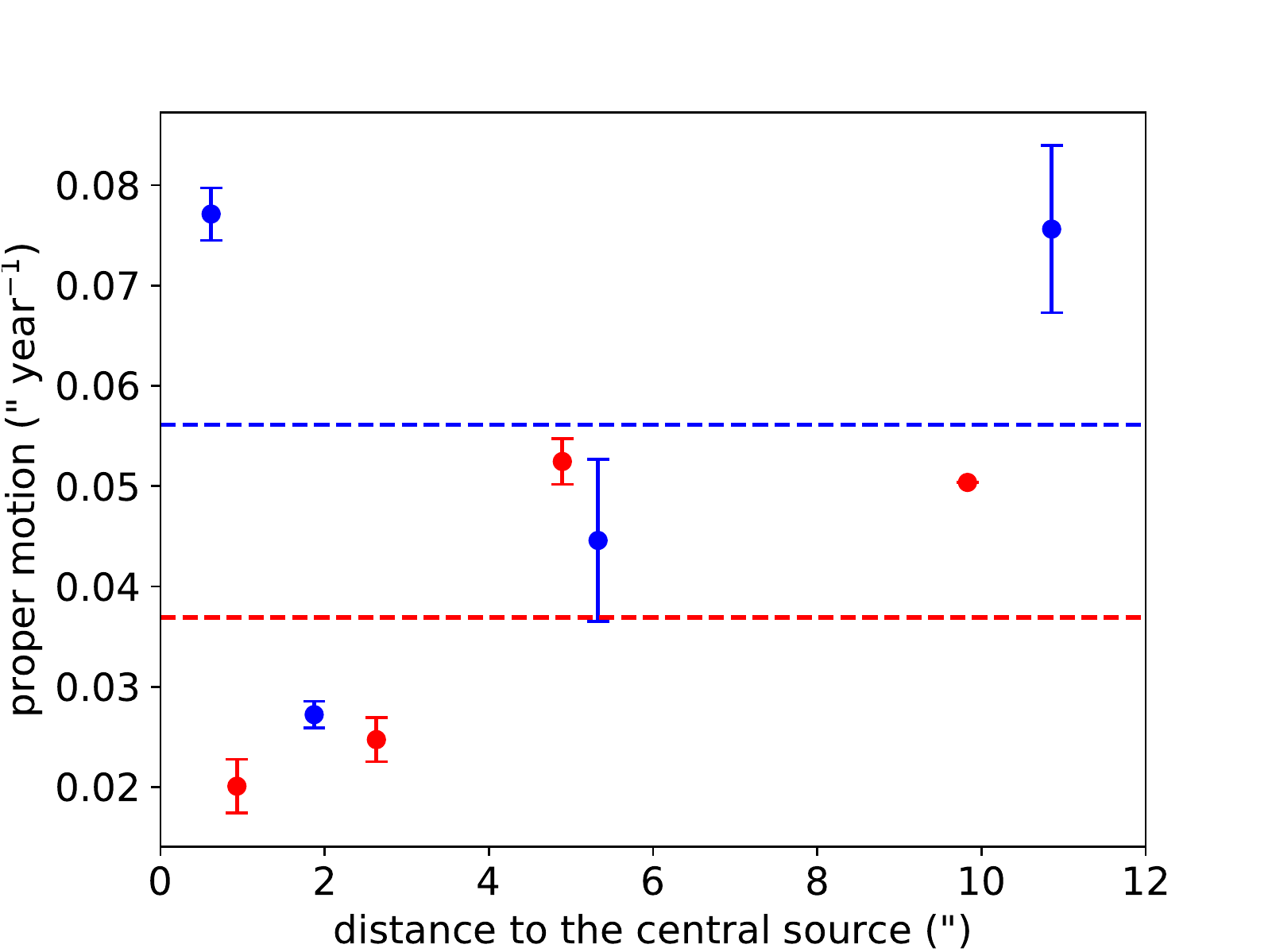}
\caption{Proper motions of the knots and the distance from the central source. The broken lines show the averaged value representing the red- and blue-shifted sides of the jet.\label{fig:ppk}}
\end{figure}

As a result, in Figure \ref{fig:comp}, the slopes of the lines that connect the knots in three epochs reflect the proper motion of the knots.
Assuming that each knot moves at a constant velocity, we performed a linear fitting in the space domain.
Figure \ref{fig:ppk} shows a plot of the derived proper motions of the knots versus the distances of the knots from the central source.
The uncertainty comes from the linear fitting given the errors of the positions derived from the 2D-hyperboloid fitting.
The averages of the proper motions are $\sim0\farcs06~{\rm yr^{-1}}$ for the blueshifted side and $\sim$0$\farcs04$ yr$^{-1}$ for the redshifted side.
The corresponding transverse velocities are $\sim78~\kms$ and $\sim52~\kms$ for the blue- and red-shifted jet, respectively.

\begin{figure}
\epsscale{1.2}
\plotone{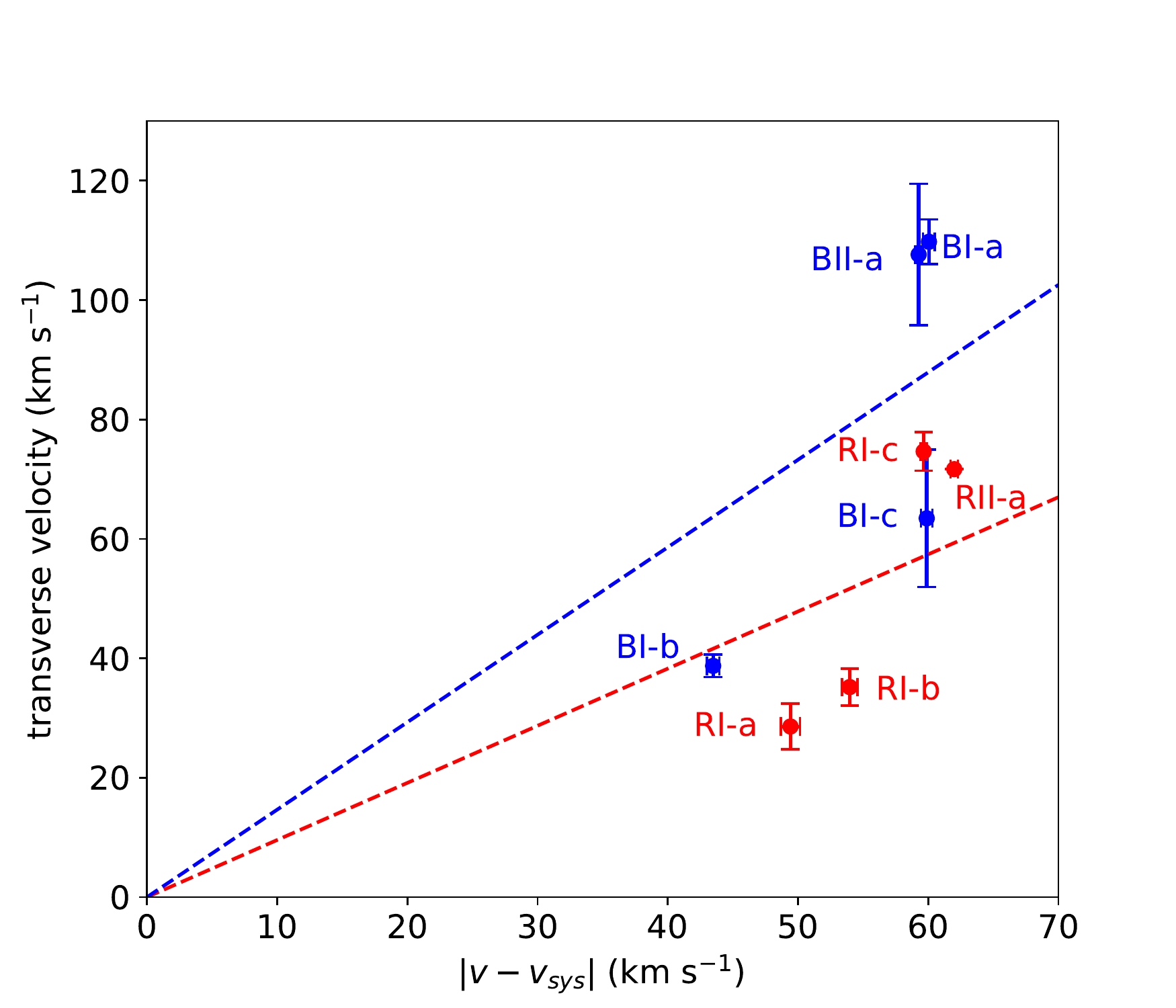}
\caption{Radial versus transverse velocities of each knot. The uncertainty of the transverse velocity originates from the linear fitting given the position errors of the 2D-hyperboloid fitting, and that of the radial velocity comes from the error of the 2D-hyperboloid fitting itself. The dashed lines mean the fitted line for the knots on each side. The angles from the the plain of the sky (y-axis) are $34 ^\circ$ for the blue line, and $46 ^\circ$ for the red line. \label{fig:vv}}
\end{figure}

The transverse versus radial velocities of the knots are plotted in Figure \ref{fig:vv}.
The dashed lines show the results of linear fitting to the knots on each side; this fitting gives the inclination angles measured with respect to the plane of the sky to be $(34 \pm 4) ^\circ$ and $(46 \pm 5) ^\circ$ for the blue- and red-shifted jet, respectively.
The derived inclination angles are consistent with those estimated through the model of infrared scattered light from the outflow cavity \citep{tob07}.
They are also consistent with the inclination angle at the base of the redshifted jet derived from the proper motions of the H$_2$O maser spots \citewt{hir11}{$\sim$40$^{\circ}$ at a distance of 293 pc,}.
Besides, the 3D velocity of each jet knot was measured by using its transverse and radial velocities.
The representative 3D velocities of the jets ($\sim98\pm4~\kms$ and $\sim78\pm1~\kms$ for the blue- and red-shifted side, respectively) were derived by averaging that of every knot on each side.

\begin{figure}
\epsscale{1.2}
\plotone{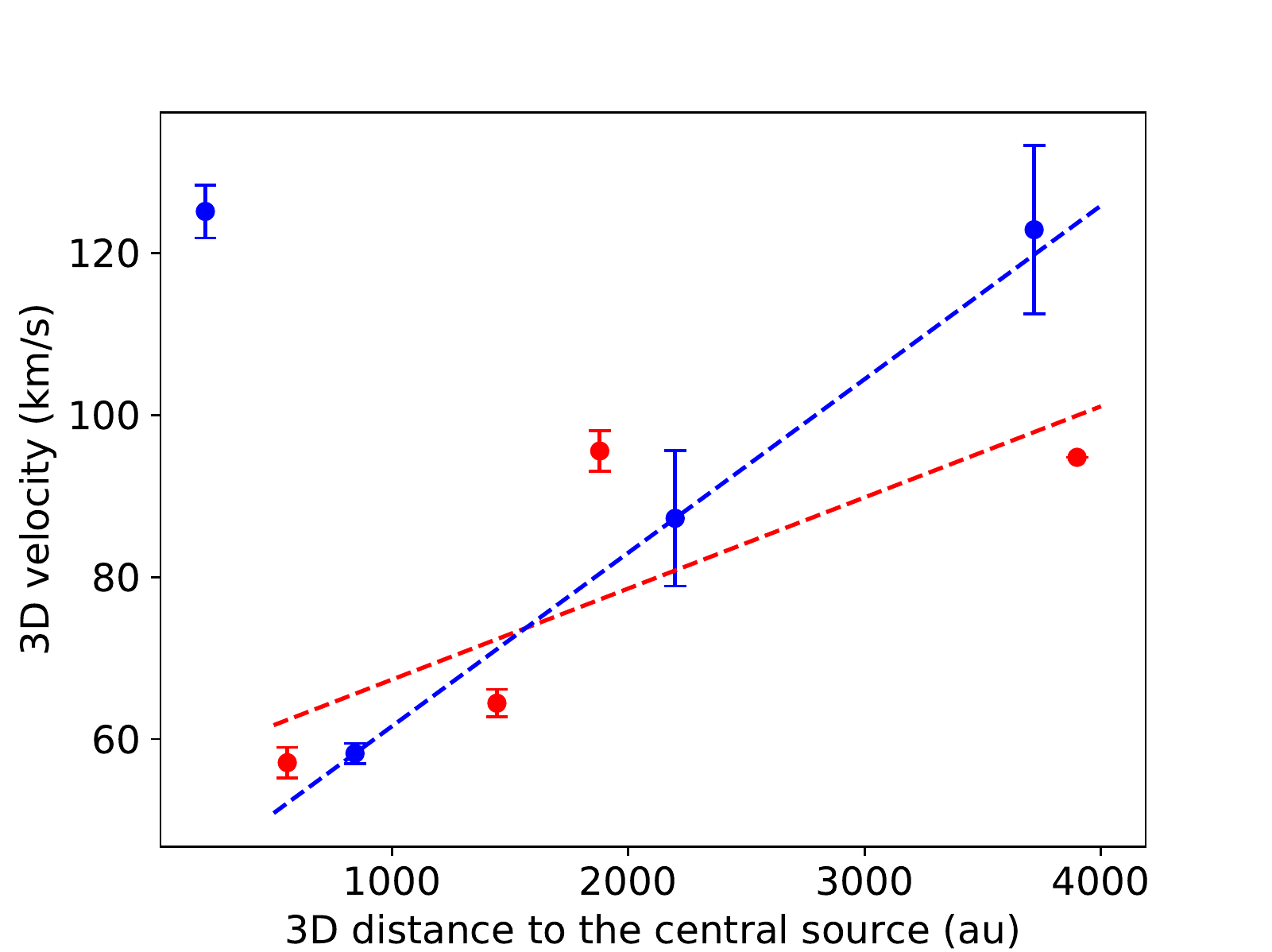}
\caption{3D velocities of each knot versus its 3D distance from the central star in the first epoch. The dashed lines indicate the linear fits to the data points except for the BI-a knot. \label{fig:hubble}}
\end{figure}

Figure \ref{fig:hubble} shows the 3D velocity versus the 3D distance to the central source in the first epoch; the 3D distance here is measured by using the inclination angles.
The 3D velocity of the knot tends to increase as the distance increases. 
This apparent acceleration seen in Figure \ref{fig:hubble} can be caused by decreasing density along the jet.
Indeed, the opening angle of $\sim10^\circ$ (Figure \ref{fig:mom0}) of the jet suggests that the jet decreases its density as it travels.
Another possibility is the variation in the ejection velocity of the jet itself.
If the ejection velocity gradually decreases as a function of time, the measured velocity becomes larger as the distance from the central source increases.
In fact, it is clear that the newly ejected BI-a knot has a higher velocity than that of BI-b, suggesting that the ejection velocity can vary.
The decrease/increase of ejection velocity can be interpreted as a consequence of expansion/contraction of the jet launching radius possibly due to the variation in mass-accretion rate.

In addition, the BII-a knot is $\sim50\%$ faster than the RII-a knot.
Such asymmetry in jet velocities are often seen in the optical/infrared jets from Class II sources \citeg{hir94}. Theoretically, it is proposed that the asymmetry occurs in the
MHD disk winds if each side of the disk has different magnetic lever arm and/or launch radius \citep{fer06}. 
\citet{dyd15} argued that the magnetic field lines in the dense disk oscillate periodically, producing the field line in one side to be more perpendicular to the disk
(reducing outflow) and another side more inclined (favor to produce outflow).
Another possibility is that the asymmetry is a consequence of different mass loading on the opposite sides of the disk in the framework of X-wind model \citep{liu12}. 
Such a velocity asymmetry has also been observed in the jet of Class 0 source L1157-mm \citep{pod16}, which suggests that the mechanism similar to the asymmetric jets in Class II sources also works in the Class 0 cases.

\begin{center}
\begin{table*}

\caption{Properties of knots}
\label{tab:knots}

\begin{center}
\begin{tabular}{cccccccccccc}
\hline \hline
 & \multicolumn{2}{c}{BI-a} & & \multicolumn{2}{c}{BI-b} & & \multicolumn{2}{c}{BI-c} & & \multicolumn{2}{c}{BII-a} \\ \cline{2-3} \cline{5-6} \cline{8-9} \cline{11-12}
 & Offset & $|v-v_{\rm sys}|$ & & Offset & $|v-v_{\rm sys}|$ & & Offset & $|v-v_{\rm sys}|$ & & Offset & $|v-v_{\rm sys}|$ \\ 
 & (arcsec) & (km s$^{-1}$) & & (arcsec) & (km s$^{-1}$) & & (arcsec) & (km s$^{-1}$) & & (arcsec) & (km s$^{-1}$) \\ 
 \hline
2006 & 0.62 & 66.0 & & 1.87 & 44.7 & & 5.33 & 59.0 & & 10.9 & 59.8 \\
2010 & 0.94 & 58.2 & & 1.93 & 42.3 & & 5.36 & 59.5 & & 11.0 & 59.7 \\
2017 & 1.45 & 55.9 & & 2.18 & 43.4 & & 5.86 & 61.2 & & 11.7 & 58.3 \\ \hline \hline
 & \multicolumn{2}{c}{RI-a} & & \multicolumn{2}{c}{RI-b} & & \multicolumn{2}{c}{RI-c} & & \multicolumn{2}{c}{RII-a} \\ \cline{2-3} \cline{5-6} \cline{8-9} \cline{11-12}
& Offset & $|v-v_{\rm sys}|$ & & Offset & $|v-v_{\rm sys}|$ & & Offset & $|v-v_{\rm sys}|$ & & Offset & $|v-v_{\rm sys}|$ \\ 
 & (arcsec) & (km s$^{-1}$) & & (arcsec) & (km s$^{-1}$) & & (arcsec) & (km s$^{-1}$) & & (arcsec) & (km s$^{-1}$) \\ 
 \hline
2006 & 0.93 & 49.5 & & 2.63 & 53.6 & & 4.89 & 59.2 & & 9.83 & 62.6 \\
2010 & 1.02 & 50.0 & & 2.71 & 54.6 & & 5.03 & 60.3 & & 10.0 & 60.8 \\
2017 & 1.12 & 48.7 & & 2.96 & 53.7 & & 5.48 & 59.5 & & 10.4 & 62.6 \\ \hline
\end{tabular}
\end{center}
\end{table*}
\end{center}

\subsection{Physical parameters of the EHV jets}
\label{sec:physpara}

Because of the unknown SiO abundance, the physical parameters of the EHV jet were estimated using the CO flux measured in the velocity ranges of $\Delta V=\pm41–70~\kms$ from the systemic velocity ($v_{\rm LSR} \sim 5~\kms$).
In the velocity ranges of $\Delta V=\pm41-50~\kms$, the CO emission from the jets is contaminated by the outflow shells associated with L1448C(N) (see Section \ref{sec:physpara_out}). Thus, in order to exclude the contamination of the shell components, we narrowed the region in the ranges of $\Delta V=\pm41-50~\kms$. We used the image observed in 2006 with the compact and extended configurations, because it recovered 80--100\% of the CO flux at $\Delta V > \pm$20 km s$^{-1}$.
On the other hand, the CO images observed in 2010 and 2017 without compact configuration suffer from the missing flux of up to $90$\% as compared to previous single-dish observations \citep{nis00} because the CO emission is spatially extended even in the EHV velocity ranges.
We assumed that the CO emission from the jet is optically thin, and that the excitation condition of the CO follows the local thermal equilibrium (LTE). The mean atomic weight was adopted to be 1.41. The CO abundance of 4$\times$10$^{-4}$ is assumed. This high value is proposed by the chemical model of protostellar winds in \citet{gla91}.
We assume an excitation temperature of 100 K as used in \citet{hir10} for comparison.

\begin{deluxetable}{lcc}
\tablecolumns{3}
\tablewidth{0pc}
\tablecaption{Dynamical parameters of the EHV jet from L1448C(N)}
\tablehead{
\colhead{Parameters} & \colhead{Blue} & \colhead{Red}}
\startdata
Mass\tablenotemark{a} ($M_{\odot}$) & 4.3$\times$10$^{-4}$ & 3.7$\times$10$^{-4}$ \\
Momentum\tablenotemark{a} ($M_\odot$ km s$^{-1}$) & 0.043 & 0.031\\
Kinetic energy\tablenotemark{b} (erg) & 4.4$\times$10$^{43}$ & 2.5$\times$10$^{43}$ \\
Mean velocity (km s$^{-1}$) & $98\pm4$ & $78\pm1$ \\
Dynamical timescale (yr) & 390 & 600 \\
Mass-loss rate\tablenotemark{a} ($M_\odot$ yr$^{-1}$) & 1.1$\times$10$^{-6}$ & 6.6 $\times$10$^{-7}$ \\
Mechanical power\tablenotemark{b} ($L_\odot$) & 0.93 & 0.34 \\
\enddata
\tablenotetext{a}{The typical uncertainty is estimated to be $\sim10~\%$.}
\tablenotetext{b}{The typical uncertainty is estimated to be $\sim20~\%$.}
\end{deluxetable}
\label{tab:jet}

The derived physical parameters of the jets are given in Table \ref{tab:jet}. The mass of the EHV jet is $\sim$8.0$\times 10^{-4}~M_\odot$.
The uncertainty of the mass estimation mainly arises from that of the absolute flux calibration. Since the SiO flux values at the middle knots of the jet measured in the three epochs show good agreement with each other (see Section \ref{sec:dimsio}), the measured flux is accurate to $\sim 10\%$, assuming that the variability of the SiO flux in the middle of the jet is negligible.
The dynamical timescale of the jets was given as $l/v$, where $l$ is the 3D length of the jet and $v$ is the mean 3D velocity of the jet.
Here, the projected length of the blueshifted CO jet was measured to be $\sim$22$\arcsec$ and, that of the redshifted CO jet was $\sim$23$\arcsec$. \citewt{hir10}{see Figure 5 in}.
The mean velocities of the jets were given in Section \ref{sec:propsio}. Thus, the dynamical timescale of the blueshifted jet and the redshifted jet were estimated to be $\sim$390 yr and $\sim$600 yr, respectively.
As a result, we obtained the total mass-loss rate of the jet to be $\sim$1.8$\times 10^{-6}~M_\odot$ yr$^{-1}$.
The total mechanical power of the jet was found to be $\sim$1.3 $L_\odot$.

\begin{figure}
\epsscale{1.0}
\plotone{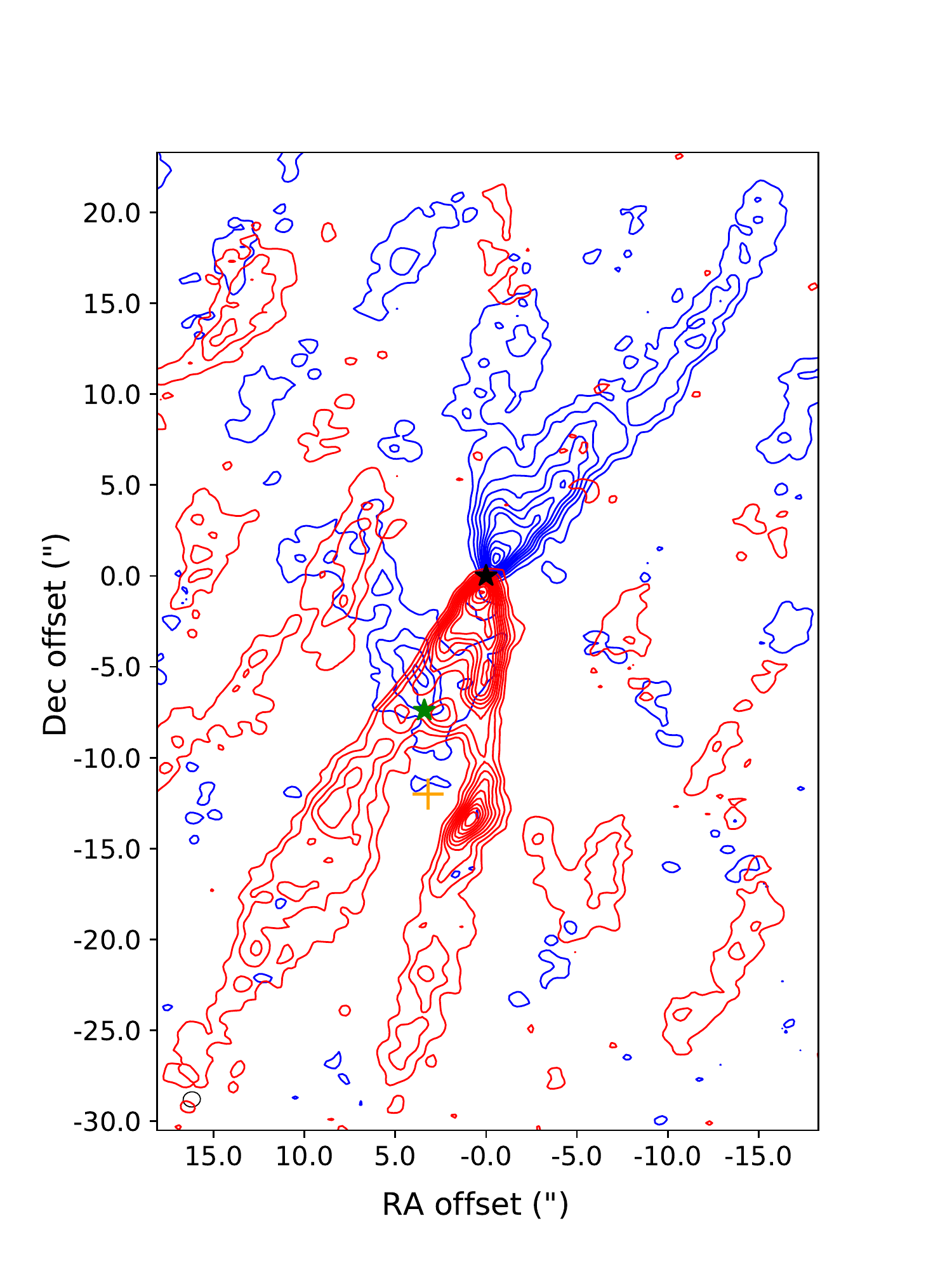}
\caption{CO $J=3-2$ map in 2006. The velocity ranges are $\pm1-40$ km s$^{-1}$ with respect to the systemic velocity. The contour interval is $\sim$2.85 $\JB$ km s$^{-1}$ (3$\sigma$) with the lowest contour level of $\sim$2.85 $\JB$ km s$^{-1}$ (3$\sigma$). The black and green stars indicate the central sources of the L1448C(N) and L1448C(S), respectively. The orange cross marks the position of RII-b knot. \label{fig:comom0}}
\end{figure}

\subsection{Physical parameters of the outflow shells}
\label{sec:physpara_out}

The CO outflow shells associated with L1448C(N) are also detected (Figure \ref{fig:comom0}). The extended shell-like structures are only seen in the velocity ranges of $\Delta V\sim \pm1-50$ km s$^{-1}$.
The physical parameters of the outflow shells were calculated using CO data in 2006.
Even with the compact configuration, the missing flux in this velocity range is significant.
By comparing to the single-dish CO $J=3-2$ data of \citet{nis00}, we estimated that the missing flux in the lowest velocity range is up to $\sim$80\%. The missing flux correction was done channel by channel.

We assumed that the CO emission is optically thin, the excitation condition of the CO follows the LTE, the CO abundance relative to ${\rm H_2}$ is 10$^{-4}$, and the mean atomic weight is 1.41. Note that we adapt the standard abundance because we consider that most of the gas in the lower velocity outflow shells is the swept-up ambient material.
We also assumed the excitation temperature of $\sim$40\ K following \citet{hir10}. It was estimated from the peak brightness temperature of CO around the systemic velocity, assuming that the CO emission at this velocity is optically thick. To calculate the dynamical parameters, the inclination angles of the outflows are assumed to be the same as those of the jets.

\begin{deluxetable*}{lccccc}
\tablecolumns{6}
\tablewidth{0pc}
\tablecaption{Dynamical parameters of the L1448C(N) outflow shell}
\tablehead{
\colhead{}&\multicolumn{2}{c}{Blue}&\colhead{}&
\multicolumn{2}{c}{Red}\\
\cline{2-3} \cline{5-6}
\colhead{Parameters} & \colhead{uncorrected\tablenotemark{a}}   & \colhead{corrected\tablenotemark{b}}    &
\colhead{}    & \colhead{uncorrected\tablenotemark{a}}   & \colhead{corrected\tablenotemark{b}}}
\startdata
Mass\tablenotemark{d} ($M_{\odot}$) & 2.6$\times$10$^{-3}$ & 9.1$\times$10$^{-3}$ & & 4.0$\times$10$^{-3}$ & 1.2$\times$10$^{-2}$ \\
Momentum\tablenotemark{d} ($M_\odot$ km s$^{-1}$) & 0.094 & 0.16 && 0.12 & 0.23\\
Kinetic energy\tablenotemark{e} (erg) & 4.9$\times$10$^{43}$ & 6.3$\times$10$^{43}$ && 5.2$\times$10$^{43}$ & 7.7$\times$10$^{43}$ \\
Momentum supply rate\tablenotemark{c,d} ($M_\odot$ km s$^{-1}$ yr$^{-1}$) & 3.1$\times10^{-5}$ & 5.3$\times$10$^{-5}$ && 2.6$\times10^{-5}$ & 5.9$\times10^{-5}$ \\
Mechanical power\tablenotemark{c,e} ($L_\odot$) & 0.13 & 0.17 && 0.14 & 0.21 \\
\enddata
\tablenotetext{a}{The effect of missing flux is not corrected.}
\tablenotetext{b}{The effect of missing flux is corrected channel by channel in comparison with the single-dish observations.}
\tablenotetext{c}{The dynamical timescale is assumed to be $\sim$3000 yr for the blueshifted jet and $\sim$4700 yr for the redshifted jet.}
\tablenotetext{d}{The typical uncertainty is estimated to be $\sim10~\%$.}
\tablenotetext{e}{The typical uncertainty is estimated to be $\sim20~\%$.}
\end{deluxetable*}
\label{tab:outf}

Table \ref{tab:outf} shows the calculated physical parameters of the outflow shells with and without missing flux correction.
After the correction, the total mass of the outflow shells was estimated to be $\sim$2.1$\times 10^{-2}~M_\odot$. The dynamical timescale of the outflow is defined as $l/v$, where $l$ and $v$ are the 3D length and the 3D velocity of the outflow. The projected length of the CO outflows on both sides was estimated to be $\sim$25$\arcsec$. Thus, the dynamical timescale of the blueshifted shell is calculated to be $\sim$3000 yr, and that of the redshifted shell to be $\sim$4700 yr. The momentum supply rate and the mechanical power of the outflow shell are now estimated to be $\sim$1.1$\times 10^{-4}~M_\odot$ km s$^{-1}$ yr$^{-1}$ and $\sim0.35~L_\odot$ with missing flux correction.
Since the mechanical power of the two-sided jet is $\sim1.3~L_\odot$, the jet has sufficient power to drive the outflow shell. 
In addition, the momentum supply rate of the jet is $\sim1.6\times10^{-4}~M_\odot$ km s$^{-1}$ yr$^{-1}$ for both sides of the jet, which is comparable to the value of the outflow shell. Therefore, the shells can be driven by the jet, although we cannot exclude the possibility that the outflow and the jet are launched independently \citep[e.g.][]{mac08,mac14}.

\subsection{Appearance of a new knot}

\begin{figure*}[]
\epsscale{1.1}
\plotone{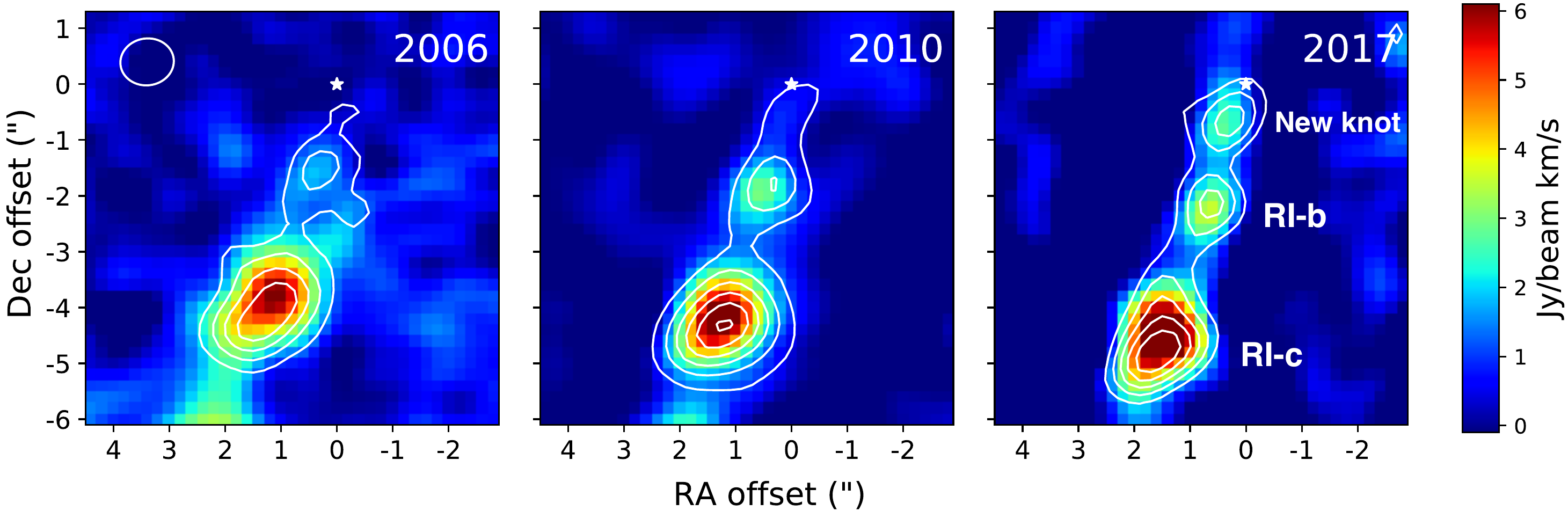}
\caption{SiO map (contours) and CO map (color scale) in three epochs. Both the SiO and CO emission were integrated over the velocity range of $\Delta V \sim 61-70~\kms$.
The contours are drawn every $\sim$1.2 $\JB \kms$ (3$\sigma$) with the lowest contour at $\sim$1.2 $\JB \kms$ (3$\sigma$). The noise level of the CO map is $\sim$0.41 $\JB \kms$. White star marks the central source. \label{fig:newknot}}
\end{figure*}

Figure \ref{fig:newknot} shows the SiO moment 0 maps in the RI region overlaid on those of CO in three epochs. The integrated ranges of these maps are $\Delta V = 61-70~\kms$ for both SiO and CO lines.
A newly appeared jet knot is seen in the 2017 image at a position near to RI-a; this component is also seen in the PV diagram of 2017 with a velocity much higher than that of RI-a (Figure \ref{fig:pv}).
The new knot is located at a distance of $\sim0\farcs77$ from the central star.
Although the new knot is clearly seen in the high-velocity moment 0 map, there is no clear boundary between it and RI-a in the PV diagram (Figure \ref{fig:pv}).
However, toward the position of the new knot, the CO/SiO intensity increases significantly in the wide velocity range of $\Delta V\sim30-70~\kms$.

We define the mass of the new knot as the mass increased between 2006 and 2017. Since this new knot is also seen in the CO map in the same velocity range, the mass of the new knot was derived using the CO map.
In order to estimate the increased mass, we firstly derived the total mass of the RI-a and RI-b knots considering the proper motions. The velocity range was taken to be $\Delta V = 51-70~\kms$.
The same physical conditions used for the EHV jet (see Section \ref{sec:physpara}) were assumed in the estimation. The total mass of the RI-a + RI-b knots is estimated to be $\sim1.5\times$10$^{-4}~M_\odot$ in 2006,  $\sim1.6\times$10$^{-4}~M_\odot$ in 2010, and $\sim1.8\times$10$^{-4}~M_\odot$ in 2017.
Then, the mass of the new knot is estimated to be $\sim$3.4$\times 10^{-5}~M_\odot$ by subtracting the average of the 2006 and 2010 values from the one of 2017. 
If we assume that the 3D velocity of this knot is same as the average velocity of the redshifted jet, $\sim$78 \ $\kms$, this knot has a momentum of $\sim$2.7$\times 10^{-3}~M_\odot$ \ $\kms$ and a mechanical power of $\sim0.85~L_\odot$.
If the new knot is ejected along the axis of the redshifted jet, the distance between this knot and the central source is $\sim$330 au.
With the 3D velocity of $\sim$78 km s$^{-1}$, the dynamical timescale of the knot is estimated to be $\sim 20$ yr.
Thus, the averaged mass-loss rate of the new knot is estimated to be $\sim 1.7 \times 10^{-6}~M_\odot~{\rm yr^{-1}}$.
However, our assumption of the excitation temperature might be underestimated in such a newly shocked region.
In addition, the dynamical timescale is the upper limit for duration time of the jet eruption. 
Therefore, the mass-loss rate derived here is considered as the lower limit.

\begin{deluxetable}{lc}
\tablecolumns{2}
\tablewidth{0pc}
\tablecaption{Dynamical parameters of the new knot}
\tablehead{\colhead{Parameters}&\colhead{the new knot}}
\startdata
Mass\tablenotemark{a} ($M_{\odot}$) & $3.4\times10^{-5}$ \\
Momentum\tablenotemark{b} ($M_\odot$ km s$^{-1}$) & $2.7\times10^{-3}$ \\
Mean velocity (km s$^{-1}$) & $78\pm1$ \\
Dynamical timescale (yr) & 20 \\
Mass-loss rate\tablenotemark{b} ($M_\odot$ yr$^{-1}$) & $1.7\times10^{-6}$ \\
Mechanical power\tablenotemark{b} ($L_\odot$) & 0.85 \\
\enddata
\tablenotetext{a}{The typical uncertainty is estimated to be $\sim10\%$.}
\tablenotetext{b}{The typical uncertainty is estimated to be $\sim20\%$.}
\end{deluxetable}
\label{tab:newknot}

\subsection{Brightness variability of the jet knots}
\label{sec:dimsio}
\begin{figure*}
\epsscale{1.1}
\plotone{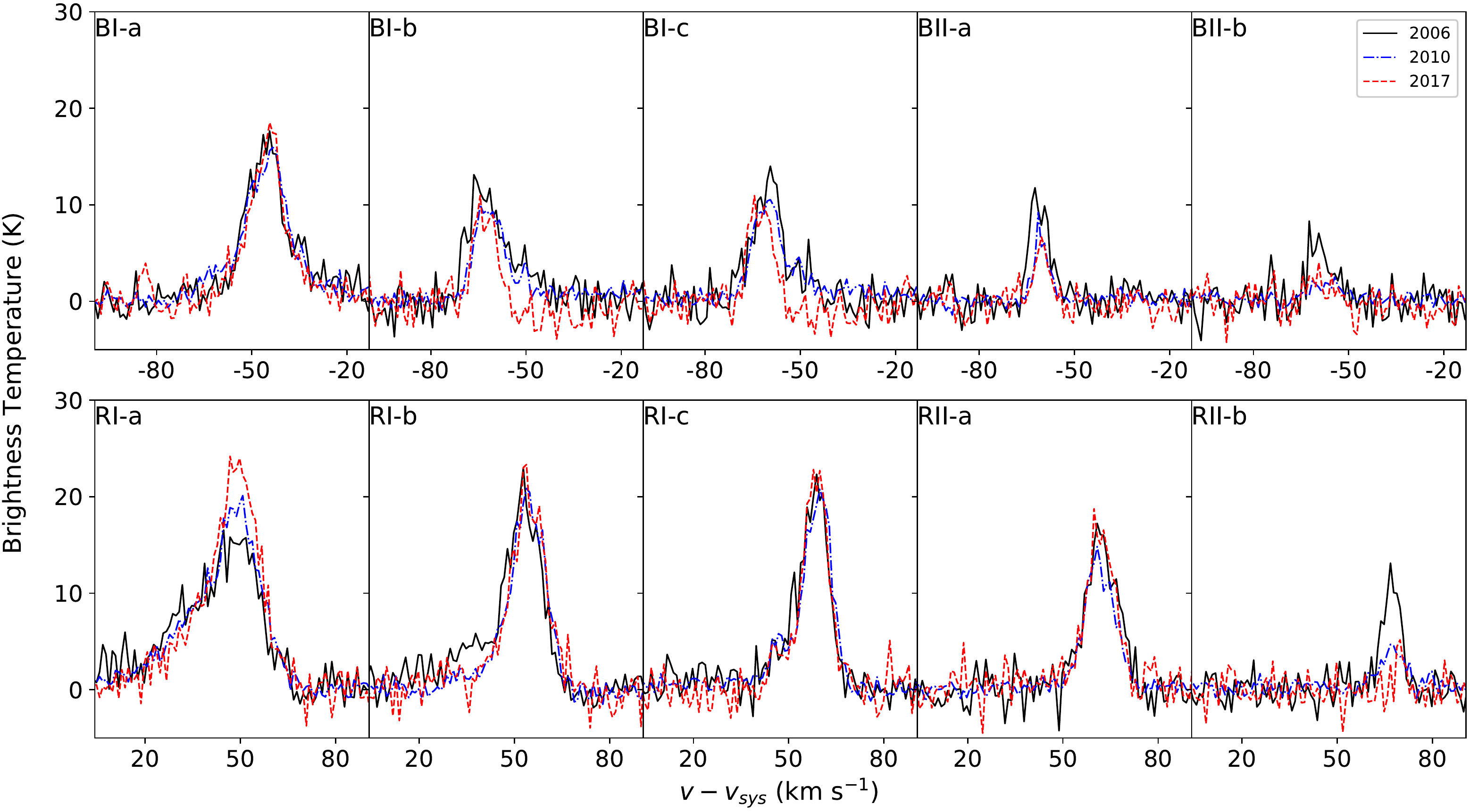}
\caption{Line profiles of SiO $J=8-7$ in the three epochs at the peak of each knot. The x-axis is the velocity ($\kms$) relative to the systemic velocity, and the y-axis is the brightness temperature (K). The black solid line, the blue dash dotted line, and the red dashed line shows the line profile in 2006, 2010 and 2017, respectively. 
\label{fig:lprof}}
\end{figure*}

\begin{figure*}[]
\epsscale{1.1}
\plotone{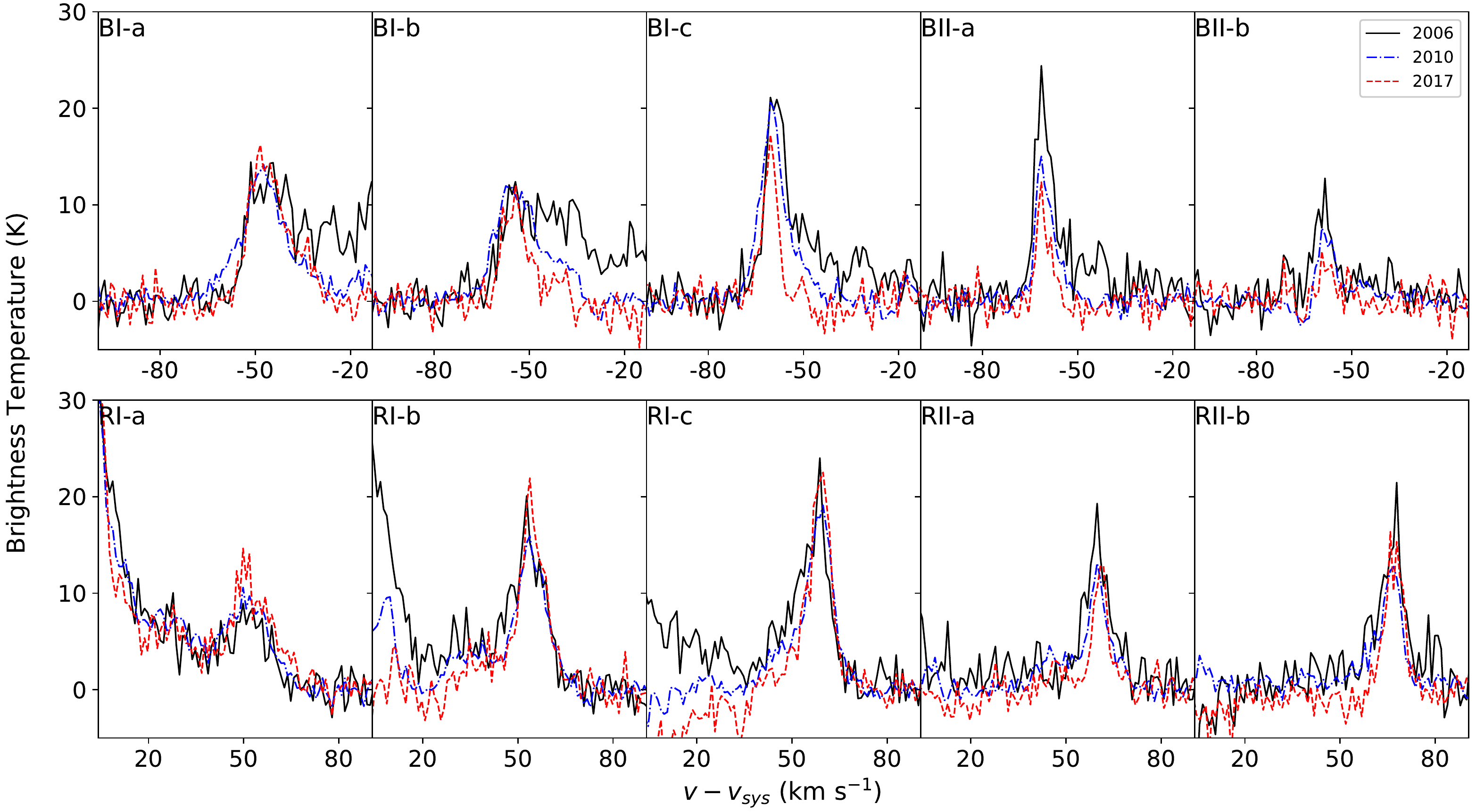}
\caption{Line profiles of CO $J=3-2$ in the three epochs at the peak of the CO emission in each knot. The x-axis is the velocity ($\kms$) relative to the systemic velocity, and the y-axis is the brightness temperature (K). The black solid line, the blue dash dotted line, and the red dashed line shows the line profile in 2006, 2010 and 2017, respectively.
\label{fig:lprofco}}
\end{figure*}

The line profiles at the emission peaks of the SiO knots are shown in Figure \ref{fig:lprof}.
The line profiles in the three epochs are consistent toward the BI-a, RI-b, RI-c, and RII-a knots.
In the BI-b and BI-c knots, the peak intensity remains similar, although the intensity of the lower velocity tail measured in 2017 is lower than those of 2006 and 2010.
The peak intensity of the RI-a knot increased significantly, from $\sim$15 K in 2006 to $\sim$25 K in 2017.
This likely originates from the appearance of the new knot.
On the other hand, the knots downstream, --- i.e. BII-b and RII-b --- have dimmed, especially RII-b; in 2006, the peak brightness temperature of the RII-b knot was $\sim$15 K, while, it became $\sim$5 K in 2017.
Since the intensity values of the knots in the middle, i.e. BI-b, BI-c, BII-a, RI-b, RI-c and RII-a, and BI-a are constant over these 3 epochs, the intensity difference in the downstream knots is not attributed to the flux calibration uncertainties.
For comparison, Figure \ref{fig:lprofco} shows the CO $J=3-2$ line profiles toward the CO peaks in the corresponding knots, although the knotty structures in CO are not as clear as those in SiO.
In the case of CO, there is no significant dimming in the RII-b knot.
Note that the separation between the SiO and CO peaks in each knot is up to $\sim1\arcsec$.

\begin{figure*}[]
\epsscale{1.1}
\plotone{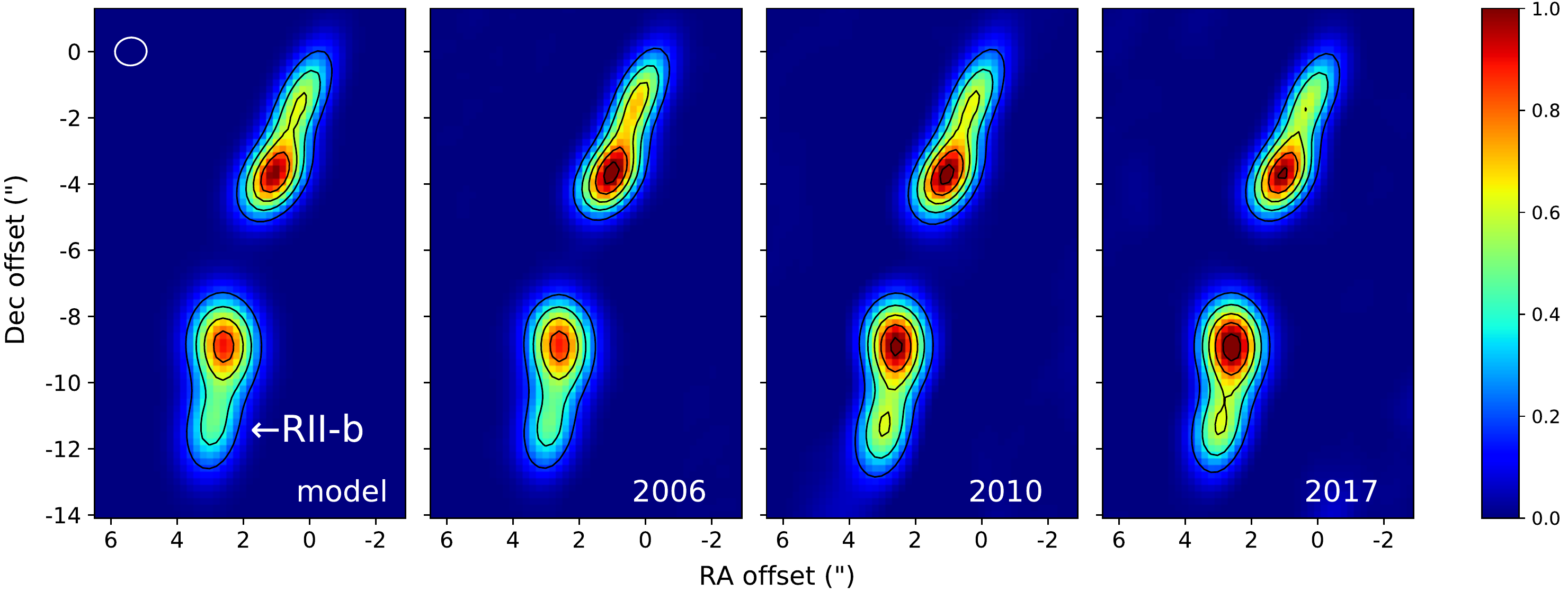}
\caption{Results of the simulated observations using the same {\it uv} coverage of the observations at each epoch. The leftmost panel shows the input model. The intensity is normalized by the maximum value in the input model. The contours start from 0.1 and drawn every 0.2.
\label{fig:sim}}
\end{figure*}

Since the observations of the three epochs have been done in different array configurations, the effect of spatial filltering to the observed intensity might be different in each epoch.
To study the effect of the spatial filtering, we have simulated the observations of the redshifted jet using the same {\it uv} sampling as that of the observations of each epoch with the MIRIAD package.
The input model was composed of four deconvolved 2D-Gaussians obtained by fitting to the SiO moment 0 map in 2006 (Figure \ref{fig:comp}).
The modeled images were convolved and re-grided to the image of the first epoch.
Simulation results (Figure \ref{fig:sim}) show that the brightness of the knots such as RI-c and RII-a is not always constant, implying that different {\it uv} sampling could introduce an intensity variation of $\sim30\%$.
On the other hand, no significant artificial dimming in the RII-b knot is produced by the different {\it uv} sampling (Figure \ref{fig:sim}).
In addition, the intensity dimming observed in the RII-b knot from $\sim15$ K to $\sim5$ K (Figure \ref{fig:lprof}) is much larger than the $\sim30\%$ variation due to the {\it uv} sampling.
Thus, we consider that the dimming of the RII-b knot in SiO should be real rather than an effect of the spatial filtering.

\subsection{Variability of the continuum emission}
\label{sec:varcon}

\begin{figure}[]
\epsscale{1.2}
\plotone{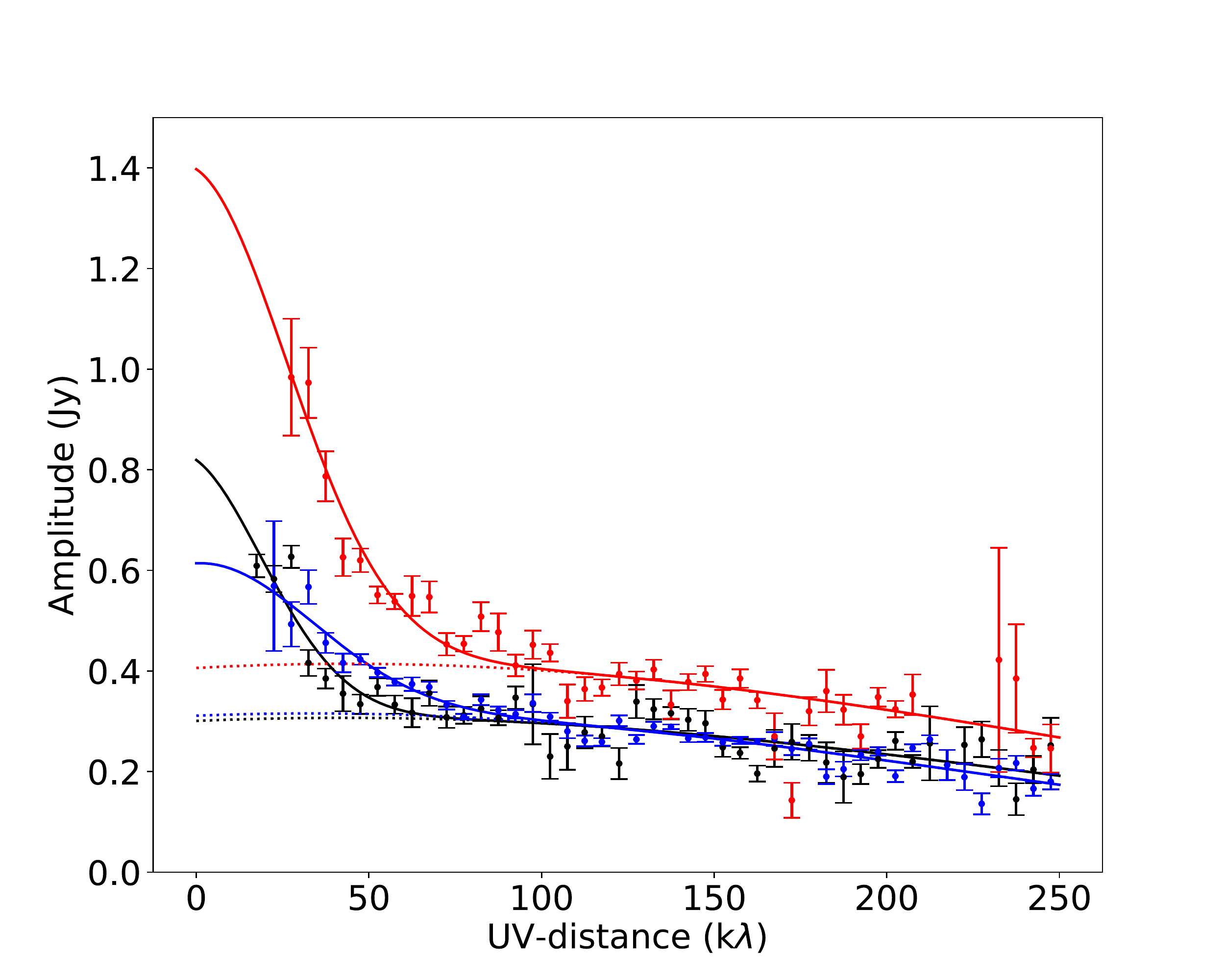}
\caption{Visibility amplitude of the continuum emission as a function of the {\it uv} distance. The black, blue, and red dots are the data points of the 2006, 2010, and 2017 observations, respectively.
The profile of each epoch was fitted with two Gaussian components.
The dotted lines indicate the Gaussian fit results to the compact component, and the solid curves are the total amplitude of the two components. \label{fig:uv}}
\end{figure}

Figure \ref{fig:uv} shows the visibility amplitudes of the continuum emission in the three epochs as a function of {\it uv} distance.
This plot was made using the sideband with the common frequency. 
Each data point was obtained by azimuthally averaging the data in a 5 k${\lambda}$ bin.
The continuum emission can be decomposed into the compact and extended components, using the boundary at $\sim70~{\rm k}\lambda$.
The visibility amplitude profile of each epoch was fit by two Gaussian components.
Due to the lack of the data points in the shortest baseline range, the Gaussian fitting is not well constrained in the extended component.
On the other hand, the missing flux does not affect the results of the compact component.
The Gaussian fit results of the compact component in the three epochs are listed in Table \ref{tab:cont}.
\begin{table}[]
\begin{center}

\caption{Gaussian fit results for the compact component}
\label{tab:cont}
\begin{tabular}{ccc}
\hline \hline
Epoch & Peak amplitude (Jy) & FWHM ($\arcsec$) \\
\hline
2006 & 0.31 & 0.37 \\
2010 & 0.32 & 0.39 \\
2017 & 0.41 & 0.35 \\
\hline
\end{tabular}

\end{center}
\end{table}
The flux values of the compact component in 2006 and 2010 are 0.30 Jy and 0.32 Jy, respectively, and are consistent with each other.
However, the flux in 2017 is 0.41 Jy and is $\sim$37\% higher than the averaged flux of 2006 and 2010, while the measured sizes of the compact component are constant over the three epochs.
Given the stability of the measured SiO flux at the middle knots (see Section \ref{sec:dimsio}), the flux calibration is considered to be reasonably good.
Such a variability in submm/mm wavelengths is also seen in the literature \citeg{liu18,yoo17}, implying a change of the mass-accretion rate.

\section{Discussion}\label{sec:dis}

\subsection{Mass-loss rate and mass-accretion rate}

The mass-loss rate by the jet is believed to be related to the mass accretion rate to the central star.
The mass-accretion rate $\dot{M}_{\rm acc}$ is expressed by
\begin{equation}
    \label{eq:mar}
    \dot{M}_{\rm acc} \sim \frac{L_{\rm bol} R_\star}{G M_\star},
\end{equation}
where $L_{\rm bol}$ is the bolometric luminosity of the central source, $R_\star$ is the stellar radius, $G$ is the gravitational constant, and $M_\star$ is the stellar mass.
The bolometric luminosity of the central source is measured to be $L_{\rm bol}\sim14~L_\odot$ (at the updated distance of 293 pc) by \citet{tob16}.
The stellar mass derived from the analysis of the Keplerian motion of the circumstellar disk was estimated to be $\sim 0.2-0.7~M_\odot$, assuming the averaged inclination angle of $\sim 40^\circ$ \citep{mar20}.
If we assume a stellar radius of $\sim3~R_\odot$ from theoretical studies \citeg{sta88}, the mass-accretion rate is estimated to be $\sim(2-6)\times10^{-6}~M_\odot~\rm{yr}^{-1}$.
If this is the case, the ratio of the mass-loss rate to the mass-accretion rate becomes $\dot{M}_{\rm loss}/\dot{M}_{\rm acc}\sim0.3-0.9$.
If we adopt a smaller stellar mass of $\sim0.2~M_\odot$, the ratio is $\sim0.3$, which is consistent with the value predicted by the X-wind model \citeg{shu00}.
Alternatively, this high $\dot{M}_{\rm loss}/\dot{M}_{\rm acc}$ implies that the averaged mass-accretion rate in the star-forming process is higher than the current value derived from the observations.
In addition, assuming a constant mass-accretion rate and stellar mass of $\sim0.2~M_\odot$, the age of the central star is estimated to be $\sim 3.1\times 10^{4}$ yr.
This value is longer than the dynamical timescale of the larger scale ($\sim 2\arcmin$) outflow \citep{bac90,gom13} by a factor of $\sim 10$. 
As a result, both the high $\dot{M}_{\rm loss}/\dot{M}_{\rm acc}$ and age of the star suggest that the mass-accretion rate and mass-ejection rate are variable.
This result is commonly found in protostellar outflows driven by low-mass YSOs. \citep{dun10,lee13,tak13,hsi16}.

\subsection{The new knot}

The mass of the new knot, $\sim$3.4$\times10^{-5}~M_\odot$, is $\sim$3 times smaller than the mean mass of the other redshifted knots, $\sim$9.3$\times10^{-5}~M_\odot$. In addition, the angular separation of $<1\arcsec$ between the new knot and the next one, RI-a, is much smaller than the typical separation of the adjacent knots.
These imply that the new knot is a ``sub-knot (smaller knotty structure)'' of RI-a.
Indeed, similar hierarchical knotty structure is also observed in the HH211 protostellar jet in the Perseus; the sub-knots in the HH211 jet have an angular separation of $\sim0\farcs75$ \citewt{lee09}{see Figure 5 in}.
The new knot in L1448C(N) could be a similar structure as the sub-knots in HH211.

The new knot has a higher velocity ($v_{\rm 3D}\sim80~\kms$) than the precedent knots.
This feature is similar to that of the BI-a knot (Figure \ref{fig:pv}), which suggests that the new knot could be the counterpart of the BI-a knot.
This is consistent with the fact that each knot likely has its counterpart (Figures \ref{fig:comp}, \ref{fig:pv} and \ref{fig:hubble}).

The appearance of the new knot suggests that the mass-accretion rate varies in a short time scale, and increased $\sim$20 yr ago (the dynamical timescale of the knot). The averaged mass-loss rate of the new knot of $\sim1.7\times 10^{-6}~M_\odot~{\rm yr}^{-1}$ is $\sim$3 times higher than the value derived from the whole redshifted jet. 

Since an increase in the mass-loss rate implies a higher mass-accretion rate, an increase of the accretion luminosity is also expected.
Assuming that the ratio of the mass-loss to mass-accretion rate, stellar radius and mass are constant, an increase of the mass-loss rate by a factor of $\sim3$ also increases the accretion luminosity by a factor of $\sim3$.
Such a luminosity change of the central source warms the circumstellar disk by $T\propto L^{0.25}$ for a blackbody \citewt{yoo17}{see}.
If this is the case, the flux density at 350 GHz is expected to be enhanced by $\sim30\%$ after the bolometric luminosity increases by a factor of $\sim3$.
Indeed, as mentioned in Section \ref{sec:varcon}, the continuum flux has increased by $\sim37\%$ between 2010 and 2017.
Taking into account the dynamical timescale of the new knot of $\sim20~{\rm yr}$, this continuum enhancement is unlikely to be the origin of the new knot. However, the enhancement in both mass-loss rate and continuum flux implies that the mass accretion in L1448C(N) varies in a short timescale of $\sim10-20$ yr.

\subsection{Dimming of the knots in the downstream}

In order to explain the dimming of the RII-b knot in the downstream of the redshifted jet, we examine the possibility of changes in density and temperature due to the knot expansion (Figure \ref{fig:lprof}).
We adopt the initial H$_2$ density of $\sim 9.0\times 10^{5}~{\rm cm^{-3}}$ and kinetic temperature of $\sim400~{\rm K}$, which are obtained from the multi-transition SiO (from $J=2-1$ to $J=11-10$) single-dish observations of \citet{nis07}.
Using the non-LTE radiative transfer code RADEX \citep{van07}, we find that an SiO column density of $\sim3.5\times10^{14}~{\rm cm^{-2}}$ is required to reproduce the peak brightness temperature of $\sim15$ K of RII-b (Figure \ref{fig:lprof}) given the line width of $\Delta V \sim 10~\kms$.
We then assume an isotropical adiabatic expansion of the knot that increases the volume by a factor of $\sim 2$.
After the expansion, the gas temperature, the H$_2$ density, and the SiO column density become $\sim270~{\rm K}$, $\sim5.0\times 10^{5}~ {\rm cm^{-3}}$, and $\sim 2.4\times10^{14}~{\rm cm^{-2}}$, respectively.
This results in a brightness temperature of SiO down to $\sim 4.7~{\rm K}$.
If this is the case, the velocity of expansion is estimated to be $\sim 30~\kms$, assuming the knot was initially concentrated within $\sim1\arcsec$ (Figure \ref{fig:comp}).
This velocity is broadly consistent with the line width of the knot in SiO/CO within a factor of $\sim2-3$ (Figures \ref{fig:lprof} and \ref{fig:lprofco}).
In this case, the change of density does not significantly affect the CO brightness temperature because of its low critical density of $\sim5\times10^{3}~{\rm cm^{-3}}$ at 300 K \citep{yan10}.

However, it is unclear why only the RII-b knot is dimming, but the other knots are constant.
A possible hypothesis is that the RII-b knot is affected by the outflow from another protostar, L1448C(S) (Figure \ref{fig:mom0}).
Indeed, \citet{hir10} found that the outflows driven by L1448C(N) and L1448C(S) may interact with each other (Figure \ref{fig:comom0}) in the redshifted side.
This implies that the redshifted jet from the L1448C(N) also can be affected by the outflow from L1448C(S).
Since the RII-b knot is located near the outflow cavity of L1448C(S) on the plane of the sky (Figure \ref{fig:comom0}), the expansion of the RII-b knot can be caused by the interaction with the outflow from L1448C(S).

Another possibility is the change of ejection direction.
As seen in Figures \ref{fig:mom0} and \ref{fig:comp}, the jet is deflected at the position in between RI and RII, and BI and BII.
This deflection suggests a change of ejection direction caused by the precession or wobbling of the circumstellar disk.
In such a case, the RII-b knot is traveling in a low-density region dissipated by the former jet. On the contrary, the RII-a knot may be plowing into the fresh denser material.
If this is the case, the RII-b knot can expand easier than the knots in the current jet axis.

In addition, the SiO abundance in the knot can be variable. Assuming an optically thin SiO line and fixing the physical conditions in the LTE, the dimming of the line suggests that the abundance of the SiO molecule was reduced by a factor of two.
The abundance of the SiO molecule is supposed to drop via reaction of ${\rm SiO + OH \rightarrow SiO_2 + H}$, since OH is abundant in the shocked gas. The rate coefficient of this reaction is proposed to be $k \sim(2 - 100) \times 10^{-12}$ cm$^3$ s$^{-1}$ \citep{let00,gus08,lan90}. Thus, the corresponding reaction timescale is estimated to be $t = 1/k \sim 300 - 16000$ yr. However, the timescale of the dimming in our observations is only $\sim10$ yr, which is much shorter than the reaction timescale. 

\section{Conclusions} \label{sec:con}

We compared 3 epochs---2006, 2010 and 2017---of SMA observations to study L1448C(N) protostellar jets in SiO $J=8-7$ line and CO $J=3-2$ line. Our main conclusions are as follows.

\begin{enumerate}
    \item The proper motions of the SiO knots are measured to be $\sim0\farcs06~{\rm yr^{-1}}$ and $\sim0\farcs04~{\rm yr^{-1}}$ for the blue- and red-shifted side, respectively. The corresponding transverse velocities are $\sim78~\kms$ (blue) and $\sim52~\kms$ (red).
    The inclination angle of the EHV jet was estimated by comparing the transverse and radial velocities of the jet knots. The inclination angle (from the plane of the sky) of the blueshifted jet is $\sim(34 \pm 4) ^\circ$ and that of the redshifted jet is $\sim(46 \pm 5) ^\circ$. The jet axis is $\sim20^\circ$ more inclined than the previous estimation. The mean velocity of the EHV jet is estimated to be $\sim(98\pm4)~\kms$ for the blueshifted jet and $\sim(78\pm1)~\kms$ for the redshifted jet.
    \item The mass-loss rate of L1448C(N) is estimated to be $\sim1.8\times 10^{-6}~M_\odot$ yr$^{-1}$. The ratio of the mass-loss rate to the mass-accretion rate is found to be $\sim0.3-0.9$, depending on the current stellar mass. The lower-end value is broadly consistent with the theoretical predictions. 
    The higher-end is commonly found in protostellar jets/outflows and is interpreted by a variation in mass-accretion rate.
    The mechanical power of L1448C(N) is refined to be $\sim1.3~L_\odot$.
    \item A new knot is found in 2017 at the base of the redshifted jet. The mass of the new knot is $\sim3.4\times10^{-5}~M_\odot$. The lower limit of the mass-loss rate of the new knot is estimated to be $\sim 2.1 \times 10^{-6}~M_\odot$ yr$^{-1}$. If this is the case, the mass-accretion rate becomes $\sim2.1\times10^{-5}~M_\odot$, which is higher than the  averaged mass-accretion rate by a factor of $\sim3$.
    \item The continuum flux from the central compact source in 2017 is $\sim$37\% higher than that in 2006 and 2010. The enhancement in both mass-loss rate and continuum flux implies that the mass accretion in L1448C(N) varies in a short timescale of $\sim10-20$ yr.
    \item The SiO flux of the RII-b knot in the downstream of the redshifted jet decreased by $\sim50$\% in the last $\sim10$ yr. This dimming of SiO can be explained by the change of physical condition caused by the expansion of the knot.
\end{enumerate}


\acknowledgments
We would like to thank the SMA staff for their help during these observations.
The SMA is a joint project between the Smithsonian Astrophysical Observatory and the Academia Sinica Institute of Astronomy and Astrophysics and is funded by the Smithsonian Institution and the Academia Sinica.
We are grateful to our referee, Dr. Rafael Bachiller, for the helpful comments.
We also thank Dr. Anthony Moraghan for carefully proofreading the manuscript.
N.H. acknowledges a grant from the Ministry of Science and Technology (MoST) of Taiwan (MoST 108-2112-M-001-017, MoST 109-2112-M-001-023).


%

\facility{SMA \citep{ho04}}

\vspace{5mm}


\software{MIR \citep{qi05}, CASA \citep{mcm07}, MIRIAD \citep{sau95}, , RADEX \citep{van07}, Astropy \citep{ast13, ast18}, APLpy \citep{rob12}}

\bibliography{bib}



\end{document}